\documentclass{aa}
\usepackage{amsmath, amssymb, graphicx, natbib}
\bibpunct{(}{)}{;}{a}{}{,}
\usepackage{txfonts}
\begin{document}

\title{Density of neutral interstellar hydrogen at the termination shock from Ulysses pickup ion observations}
\author{M. Bzowski \inst{1} \and E. M{\"o}bius \inst{2} \and S. Tarnopolski \inst{1} \and V. Izmodenov \inst{3} \and G. Gloeckler \inst{4}} 

\offprints{M.~Bzowski (\email{bzowski@cbk.waw.pl})}

\institute{
Space Research Centre, Polish Academy of Sciences, Bartycka 18A, 00-716 Warsaw, Poland \and
Space Science Center and Department of Physics, University of New Hampshire \and
Moscow State University and Space Research Institute RAS, Russia \and
Department of Atmospheric, Oceanic, and Space Sciences, University of Michigan}

\date{}

\abstract
    {} 
   {By reevaluating a 13-month stretch of Ulysses SWICS H pickup ion measurements near 5~AU close to the ecliptic right after the previous solar minimum, this paper presents a determination of the neutral interstellar H density at the solar wind termination shock and implications for the density and ionization degree of hydrogen in the LIC.}
   {The density of neutral interstellar hydrogen at the termination shock was determined from the local pickup ion production rate as obtained close to the cut-off in the distribution function at aphelion of Ulysses. As shown in an analytical treatment for the upwind axis and through kinetic modeling of the pickup ion production rate at the observer location, with variations in the ionization rate, radiation pressure, and the modeling of the particle behavior, this analysis turns out to be very robust against uncertainties in these parameters and the modeling. 
}
   {Analysis using current heliospheric parameters yields the H density at the termination shock equal to $0.087\pm0.022$~cm$^{-3}$, including observational and modeling uncertainties. }
   {}

\titlerunning{Density of neutral H from Ulysses pickup ion observations }
\authorrunning{Bzowski et al.}

\maketitle

\section{Introduction}
Neutral interstellar gas of the local interstellar cloud (LIC) penetrates into the inner heliosphere as a neutral wind due to the relative motion between the Sun and the LIC. Apparently, the Sun is found near the boundary of a warm, relatively dilute cloud of interstellar gas, possibly with a significant gradient in the ionization fraction of H and He \citep[e.g.][]{cheng_bruhweiler:90a, wolff_etal:99, slavin_frisch:02} within a very structured surrounding \citep[e.g. reviews by][]{cox_reynolds:87a, frisch:95a}. In a companion paper within this special section, Frisch and Slavin (2008) lay out how the physical parameters and composition of the LIC at the location of the Sun, as derived from in-situ observations and from absorption line measurements, constrain the ionization state and radiation environment of the LIC. In situ observations of the two main constituents of the LIC, H and He, have been obtained with increasing accuracy, starting with the analysis of backscattered solar Lyman-$\alpha$ intensity sky maps \citep{bertaux_blamont:71, thomas_krassa:71} for H as well as with rocket-borne \citep{paresce_etal:74a} and satellite-borne \citep{weller_meier:74} observations of interstellar He using the solar He I 58.4 nm line. The optical diagnostics was followed by discovery of pickup ions for He \citep{mobius_etal:85a} and for H \citep{gloeckler_etal:92} and finally by direct neutral He observations \citep{witte_etal:93}.

Such in-situ diagnostics, even at 1~AU, is made possible by the neutral gas flow deep into the inner heliosphere. Through the interplay between this wind, the ionization of the neutrals upon their approach to the Sun, and the Sun's gravitational field (distinctly modified by radiation pressure for H) a characteristic flow pattern and density structure is formed, with a cavity close to the Sun and gravitational focusing on the downwind side (for all species except H). The basic understanding of the related heliosphere -- LIC interaction has been summarized in early reviews by \citet{axford:72, fahr:74, holzer:77, thomas:78}. While He provides us with almost completely unbiased information about the physical parameters of the LIC since it enters the heliosphere unimpeded, the abundance of H and O, along with other species, is significantly depleted, their speed decreased, and their temperature increased through charge exchange in the heliospheric interface \citep{fahr:91, rucinski_etal:93a, izmodenov_etal:99a, mueller_etal:00, izmodenov_etal:04a}. A consolidation of the physical parameters of interstellar He, including the flow velocity vector relative to the Sun, as determined from neutral gas, pickup ion, and UV backscattering observations, was achieved through the effort of an ISSI Team \citep[][and references therein]{mobius_etal:04a}, thus leading to a benchmark for the physical parameters of the LIC. This paper is part of a follow-up effort within an ISSI Team to also consolidate the determination of the LIC H density. The determination of the H density in the LIC proper not only involves a measurement inside the heliosphere, but is also dependent on the filtration of H in the heliospheric boundary. Therefore, consolidating the observational results concentrates on the determination of the H density at the termination shock, which still requires taking into account of the dynamics of the flow into the inner heliosphere as well as ionization effects. This paper deals with the determination of the H density from the pickup ion observations made with Ulysses SWICS. The effort to obtain the density from mass-loading of the solar wind by H pickup ions and its resulting slowdown at large distances from the Sun is described in the paper by \citet[this volume]{richardson_etal:08a}, while \citet[this volume]{pryor_etal:08a} discuss a determination of the H density based on the reduction of the modulation of the UV backscatter signal with distance from the Sun. To illustrate the state of the model-dependent H density value in the LIC \citet[this volume]{mueller_etal:08a} compare different global models of the heliosphere and their results for the distances of the key boundary structures and the filtration factor, which also connects the inner heliosphere observations to the ionization state of the LIC, discussed by \citet[this volume]{slavin_frisch:08a}.

In their previous work \citet{gloeckler_geiss:01b} used pickup ion fluxes as observed at 5~AU with Ulysses SWICS, the charge exchange rates from SWOOPS, and a Vasyliunas \& Siscoe distribution function to deduce the local neutral H density; they then used a hot interstellar gas model with the ionization rate significantly modified by electron ionization to deduce the density at the termination shock. In the present paper we use the same data set, but follow a complementary approach. 

After discussing previous derivations of the local neutral gas density and its extrapolation to the termination shock at the beginning of section 2 we present an alternative approach. We make use of the fact that the ionization rate in the pickup ion production appears both as production rate of PUI and as loss rate of the parent neutral gas population and any variations balance close to the aphelion of Ulysses. As a consequence, the PUI production rate is almost exactly proportional to the H density at the termination shock. In the same section we illustrate this behavior in a simplified analytical model that applies to the upwind region.

In section 3 we simulate the local PUI production rate, starting with the density in the interstellar medium, compare it with the observations, and confirm the robustness of this approach by varying the parameters. We start with Monte-Carlo simulations of the flow through the heliospheric interface for two different LIC parameter sets, which result in two different H densities at the nose of the termination shock. In a second step, we hand these results over to a 3D time dependent test-particle code to calculate the H densities and H$^+$ PUI production rates at Ulysses during the observation interval, while accurately taking into account losses and radiation pressure along the trajectories of interstellar gas. We find the density at the termination shock and the LIC parameters that fit the observations best by interpolating between the two initial models. We study the response of the resulting PUI production rates to variations in the ionization rate, radiation pressure, and details in the modeling. In section 4, we present the results and show that our method is very robust against uncertainties of these parameters, and, in fact, against details of the simulations.
\begin{figure}
\resizebox{\hsize}{!}{\includegraphics[width=8cm]{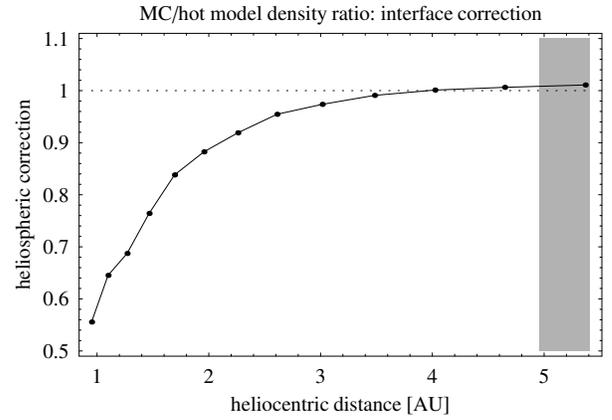}}

\caption{The interface correction: ratio of results of the Moscow Monte Carlo model for the geometry of Ulysses H PUI observations to the results of a sum of classical hot models (two populations), evaluated for identical parameters of the gas at the nose of the termination shock and identical ionization rate and radiation pressure as used in the MC model. The dotted horizontal line marks the heliospheric correction value equal to 1, the shaded area corresponds to the range of Ulysses heliocentric distances during the observations.}
\label{aa}
\end{figure}
\begin{figure}
\resizebox{\hsize}{!}{\includegraphics[width=8cm]{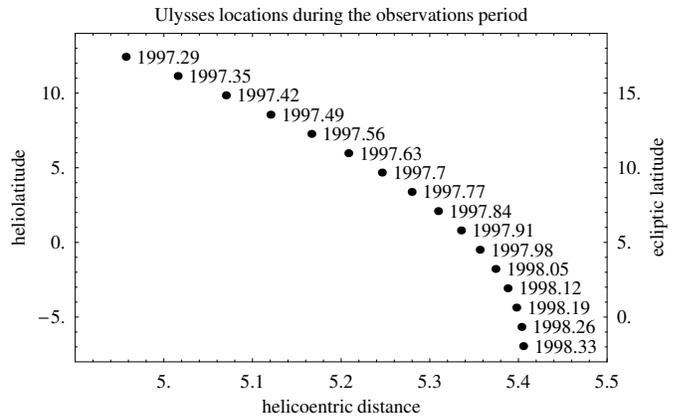}}

\caption{Change of Ulysses position during the observations. The horizontal axis shows the heliocentric distance, the left-hand vertical scale heliolatitude, and the right-hand vertical scale ecliptic latitude. The times are indicated at the plot. Ecliptic longitude varied from $153.6\degr$ at the beginning of observations interval to $157.5\degr$ at the end, which corresponds to a change from $78.4\degr$ to $81.8\degr$ in heliolongitude.
 }
\label{obs1}
\end{figure}
\begin{figure}
\resizebox{\hsize}{!}{\includegraphics[width=8cm]{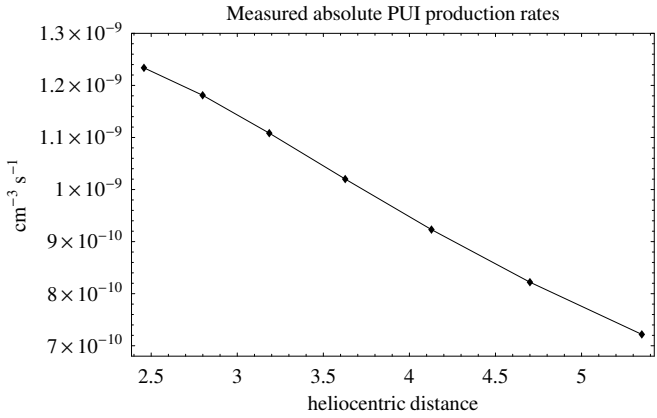}}

\caption{Absolute production rates of H PUI as a function of heliocentric distances, obtained from the observed PUI spectrum during the interval discussed in the paper. Shown are the rates after normalizing the spectrum to the production rate at 1~AU and then multiplying by $1/r^2$. }
\label{obs2}
\end{figure}

\section{Derivation of the neutral gas density from pickup ion fluxes}

The observed pickup ion flux density at any location $\vec{r}$ in the heliosphere is directly proportional to the local pickup ion source function $S\left(\vec{r}\right)$ taken just below the cut-off speed, i.e. at $w=v/v_{\mathrm{SW}}=1$, where $v$ is the pickup ion speed in the rest frame of the solar wind, that with respect to the Sun moves with $v_{\mathrm{SW}}$. The source function is given by 
\begin{equation}
\label{efMoe1}
S\left(\vec{r}\right) = \beta_{\mathrm{ion}}\left(\vec{r}\right)\, n\left(\vec{r}\right)
\end{equation}
where $\beta_{\mathrm{ion}}\left(\vec{r}\right)$ is the total local ionization rate and $n\left(\vec{r}\right)$ the local interstellar neutral gas density. Hence determining the local neutral gas density requires the knowledge of the ionization rate, and the uncertainty of the derived density is directly related to the uncertainty with which the ionization is known.

The situation appears even worse in the attempt to derive the interstellar gas density at the termination shock from observations in the inner heliosphere. For most interstellar species, except for He, ionization has already significantly depleted the local density at least where good quantitative observations of interstellar pickup ions have been available so far. In this way the local neutral gas density  is also dependent on the ionization rate, with a depletion that typically scales exponentially with the inverse of the distance from the Sun and could be written for any short local stretch as:
\begin{equation}
\label{efMoe2}
n\left(r\right) \sim \exp\left[-\frac{\alpha}{\beta_{\mathrm{ion}}\left(\vec{r}\right) r}\right]
\end{equation} 
Only the proportionality is important here for the arguments made below and not any constants, such as $\alpha$, that can be adjusted for normalization. In addition, the average ionization rate relevant for the depletion of the neutral gas (referred to as the  loss rate) may be different from that responsible for pickup ion generation (production rate) because of the different time scales involved, and, in particular, because H is also subject to radiation pressure and its variations. This combination makes any determination of the density of neutral gas at the termination shock dependent both on the modeling and on the knowledge of heliospheric parameters \citep[e.g.][]{rucinski_bzowski:96}. This is certainly true for the H PUI observations taken with Ulysses SWICS \citep{gloeckler:96a, gloeckler_geiss:01a} at or inside 5.3~AU. 

\subsection{Previous neutral gas density determination}

To minimize this influence, \citet{gloeckler_geiss:01a} used an approach that a) relies on a long-term averaging of data at Ulysses, b) uses the PUI transport model by \citet{vasyliunas_siscoe:76}, and c) simultaneously determines the total ionization rate from the slope of the pickup ion velocity distribution, which reflects the radial distribution of the neutral gas inside the observer distance. A similar approach had been taken by \citet{mobius_etal:88a} for the determination of the He density from observations at 1~AU. By making use of the fact that He$^{+}$ pickup ions are solely created by charge exchange with solar wind He$^{2+}$ and of the Ulysses SWICS capability to simultaneously observe He$^{2+}$ pickup ion and solar wind fluxes at $\sim 5$~AU, where interstellar He is not significantly depleted yet, \citet{gloeckler_etal:97} were able to obtain a He density whose uncertainty only depends on the knowledge of the charge exchange cross section and is independent of the absolute calibration of the observing instrument. For H at least two ionization processes contribute substantially, solar wind charge exchange and UV ionization, so that the rate cannot be easily eliminated from the analysis by simultaneous measurements. This remaining uncertainty in the total ionization rate, which translates into the uncertainty of the density, is exemplified by the fact that \citet{gloeckler_geiss:01b} had to invoke a rather high electron ionization rate of $2.4 \times 10^{-7}$~s$^{-1}$ at 1~AU (with a distance dependence that is stronger than $1/r^{2}$) to explain the pickup ion velocity distribution.

Recently, \citet{gloeckler_etal:08a} determined the densities of interstellar H, N, O, Ne, and Ar at the termination shock using their abundances relative to He in the energetic tails of the ion distributions in the heliosheath, as obtained with both Voyager LECP sensors with a relative uncertainty of $\pm 10$\%. To arrive at the absolute densities, they used the interstellar He density derived from Ulysses SWICS He$^{++}$ pickup ion measurements (see previous paragraph) of $0.015 \pm 0.002$~ cm$^{-3}$  \citep{gloeckler_geiss:04a, gloeckler_etal:04b}, which is also the consensus value for the interstellar He density based on these pickup ion, direct neutral gas, and UV backscattering observations \citep{mobius_etal:04a}. Combining these two observations, \citet{gloeckler_etal:08a} arrived at an H density at the termination shock of 0.08~cm$^{-3}$. With the two uncertainties cited by \citet{gloeckler_etal:08a} combined as independent contributions, the resulting uncertainty for the H density is obtained equal to $\pm 0.013$~ cm$^{-3}$, or 17\%. While this method is insensitive to uncertainties in the absolute geometric factor of the Ulysses SWICS instrument of $\pm 25$\%, cited by \citet{gloeckler_etal:08a}, which applies to the direct H pickup ion observation method, the determination of the abundances may be subject to additional uncertainties in the production rates for the different species that were used to infer the neutral species ratios.

\subsection{Alternative neutral gas density determination, minimizing the influence of uncertainties in the ionization rate}

In the following discussion we take a different approach, which will minimize the influence of uncertainties in the ionization rate on the resulting neutral gas density at the termination shock. We will illustrate this behavior in a simplified analytical treatment in this section, before we show with parameter variations in rigorous simulations, discussed in the next section, that this also holds for the actual Ulysses observations and even extends to variations in the radiation pressure. 

As becomes obvious from the relation between the neutral gas density and the pickup ion source function, a linear dependence of the resulting neutral gas density on the ionization rate (according to Eq.(\ref{efMoe1})) remains valid even at the termination shock. Conversely, an exponential behavior due to depletion by ionization (see Eq.(\ref{efMoe2})) prevails in the inner heliosphere, massively overcompensating the linear dependence of the source function on the ionization rate. Between these two extreme locations must lie a place where the effects of the ionization rate on the source function and thus on the observed pickup ion flux cancel. Here, the observed quantity is strictly proportional to the neutral density at the termination shock and -- to the first order -- does not depend on the choice of the ionization rate any more, if the adopted ionization rate is not too different from the correct value. We will explore this behavior below for inflow on the heliospheric upwind axis for which an analytical solution can be found.

In our derivation we make the reasonable assumption that radiation pressure fully compensates solar gravity: $\mu = 1$. This assumption is justified during the observation period, as discussed in detail in the Appendix. Even a significant reduction of $\mu$ has very little influence on the final result, as is shown with the simulations in section 3. For $\mu$ close to 1 the interstellar inflow speed remains almost constant as a function of distance from the Sun and equal to the interstellar bulk flow speed at the termination shock $V_{\mathrm{ISM}}$. Then the density of the neutral interstellar gas is reduced along the upwind axis according to
\begin{equation}
\label{efMoe3}
\frac{\mathrm{d} n\left(r\right)}{\mathrm{d}t} = -\beta_{\mathrm{ion}}\, n\left(\vec{r}\right).
\end{equation}
With a constant inflow speed we can use:
\begin{equation}
\label{efMoe4}
-\mathrm{d} r = V_{\mathrm{ISM}}\,\mathrm{d}t
\end{equation}
(with the gradient directed inward). The ionization rate, both solar wind charge exchange and UV combined, varies as
\begin{equation}
\label{efMoe5}
\beta_{\mathrm{ion}}\left(r\right)=\beta_{\mathrm{ion}, E}\, r_{E}^{2}/r^{2},
\end{equation}
where $\beta_{\mathrm{ion}, E}$ is the ionization rate at $r_{E}=1$~AU. Combining Eq.(\ref{efMoe3}) through Eq.(\ref{efMoe5}) together leads to
\begin{equation}
\label{efMoe6}
\frac{\mathrm{d} n\left(r\right)}{n\left(r\right)} = \frac{\beta_{\mathrm{ion}, E}\,r_{E}^2}{V_{\mathrm{ISM}}}\times \frac{\mathrm{ d} r}{r^2}.
\end{equation}
After logarithmic integration this yields:
\begin{equation}
\label{efMoe7}
\ln \frac{n\left(r\right)}{n_0} = -\frac{\beta_{\mathrm{ ion}, E}\,r_E^2}{V_{\mathrm{ ISM}}\,r},
\end{equation}
which is equivalent to
\begin{equation}
\label{efMoe7a}
n\left(r\right) = n_0\, \exp\left(-\frac{\beta_{\mathrm{ ion}, E}\,r_E^2}{V_{\mathrm{ ISM}}\,r}\right)= n_0\,\exp\left(-\frac{\Lambda}{r} \right),
\end{equation}
i.e., the density falls off exponentially with the typical penetration distance $\Lambda = \beta_{\mathrm{ ion}, E}\,r_E^2/V_{\mathrm{ ISM}}$, as the gas approaches the Sun. Consequently, the pickup ion source function is
\begin{equation}
\label{efMoe8}
S\left(r\right) = \beta_{\mathrm{ ion}, E}\,\left(\frac{r_{E}}{r}\right)^{2}\,n_0\,\exp\left[-\frac{\beta_{\mathrm{ ion}, E}\,r_E^2}{V_{\mathrm{ ISM}}\, r} \right], 
\end{equation}
which now depends linearly on the neutral density $n_0$ at the termination shock and in two ways on the ionization rate $\beta_{\mathrm{ ion}, E}$. Equation (\ref{efMoe8}) is not dependent on $\beta_{\mathrm{ ion}, E}$ anymore if ${\mathrm{d}S}/\mathrm{d}\beta_{\mathrm{ion}, E}=0$ for a given distance $r$. This condition yields
\begin{equation}
\label{efMoe9}
1 = \beta_{\mathrm{ion}, E}\,r_E^2/V_{\mathrm{ISM}}\,r
\end{equation}
or the effects of the ionization rate cancel exactly at the penetration distance $\Lambda$, i.e. at the edge of the hydrogen cavity in the heliosphere which, by definition, is the geometric location of the surface where the local density is equal to 1/e of the density at ``infinity''. For the ionization rate of $5.5\,10^{-7}$s$^{-1}$ used by \citet{gloeckler_geiss:01b} and an interstellar inflow speed of 22~km/s and temperature of $\sim 12\,000$~K at the termination shock, as results for the combined primary and secondary distributions of interstellar H from global modeling \citep{izmodenov_etal:03a} and from SOHO SWAN observations \citep{quemerais_etal:99,lallement_etal:05,costa_etal:99}, the point of perfect compensation is at 3.8~AU in the upwind direction, where the cavity edge is closest to the Sun. At crosswind, where the Ulysses observations were made, the cavity ends at $\sim 5.5$~AU, and in the downwind direction at a still larger distance. 

At these distances from the Sun the local density of neutral gas depends linearly on the local ionization rate and on the density at the termination shock $n_0$. Consequently, the source function of PUI, as defined in Eq.(\ref{efMoe1}), depends linearly on $n_0$. 

Through kinetic simulations that include all important effects in the inner heliosphere, such as ionization and radiation pressure, we will demonstrate in the following section that the Ulysses observations in 1997 through 1998  between 5 and 5.4~AU, used by \citet{gloeckler_geiss:01b} and \citet{izmodenov_etal:04a}, were indeed made in a region where the effects of a potentially not so well known ionization rate cancel and also that uncertainties in the radiation pressure as well as effects from different treatments of the particle behavior in the modeling are minimal. Henceforth, we will make use of the same data set to derive a refined value for the interstellar H density at the TS, which is robust against remaining uncertainties in the heliospheric parameters that control the local density distribution such as ionization rates and radiation pressure.
 
\section{Simulations and comparison with the observations}

In the following we will model the interstellar H distribution at the locations of the observations, starting from the pristine interstellar medium. We will reproduce the observed production rate of H pickup ions while taking into account all known and relevant heliospheric processes and their current uncertainties as much as possible. 

Calculation of the local H density at the point of Ulysses observations involves length scales that span two orders of magnitude, from the scale of the typical penetration distance of H $\Lambda \simeq 3$~AU, to the size of the heliosphere $> 100$~AU. Likewise, the time scales have a comparable range --  from weeks for solar UV illumination structures and rotation period to $\sim 40$~years of the H travel time from the pristine LIC to the observation point. To cover such ranges in a single simulation at a sufficient level of detail is beyond the reach of current computer resources available.

\subsection{Description of numerical models}

Therefore, separate simulations were performed on two different spatial scales and levels of detail, and then combined. In a first step, large scale configuration of the heliosphere was established for two sets of parameters for the LIC and solar output averaged over a large time interval, using the Moscow Monte Carlo (MC) code \citep{baranov_malama:93}. In a next step, modifications of the neutral interstellar hydrogen flow due to its interaction with the solar environment inside the termination shock were simulated using the Warsaw test particle 3D and time dependent kinetic code \citep{rucinski_bzowski:95b, bzowski_etal:97, bzowski_etal:02, tarnopolski_bzowski:07a}. 

The model adopted in the MC simulation is static and axially symmetric, assuming a constant and spherically symmetric solar particle and radiation output. The MC model is used to infer the physical parameters of H at the nose of the termination shock, i.e. the flow that will finally reach the inner solar system, where the observations are made. As extensively discussed in the past \citep{baranov_etal:98a, izmodenov:00}, neutral interstellar gas at the termination shock can be approximated to some extent by two populations (primary and secondary), featuring Maxwellian distributions shifted in velocity by specific bulk flow values. The primary population represents the original interstellar neutral H gas, while the secondary population is created between the heliopause and heliospheric bow shock due to charge exchange between interstellar atoms and the compressed and heated plasma that flows around the heliopause. These two populations are taken as boundary conditions at the termination shock for the test-particle model of the inner heliosphere.

The current Warsaw test-particle 3D and time-dependent model requires a known distribution function far away from the Sun, which is invariable in time and homogeneous in space. Consequently, the two neutral-gas populations are Maxwellians characterized by temperature, bulk velocity, and density. 

As demonstrated by \citet{baranov_etal:98a, izmodenov:01, izmodenov_etal:01a}, such assumptions are a considerable simplification. In reality, neither the distribution functions of the two populations are Maxwellian, nor their macroscopic parameters are homogeneous in space. On the other hand, it was also demonstrated \citep{izmodenov_etal:05b} that 30\% sinusoidal variations in the spherically symmetric solar wind density during the solar activity cycle induce only $\sim 10$\% variations in the thermal populations of neutral H at the termination shock. Since simulations including more realistic variations of the solar wind evolution are unavailable, and given tremendous computer burden of the MC time-dependent model, these variations were assumed to be negligible for this study. 

Deviations of the parameters of the distribution function from homogeneity at the termination shock result in systematic differences in the local densities returned by the Moscow MC and Warsaw test-particle models. To assess the robustness of our approach to such deviations, we calculated first the H density at the geometric location of the PUI data with the use of the MC model and then, adopting identical values of the radiation pressure and ionization rates and the parameters of the primary and secondary populations at the termination shock, with the Warsaw test-particle model in its axially symmetric, static mode. The ratio of the density profiles returned by the two models is shown in Fig.\ref{aa}. While the two results significantly deviate from each other closer to the Sun along the Ulysses PUI accumulation line, the MC and test particle models return almost identical results between 3.5 and 5~AU. Generally, the hot model overestimates the density by a factor which increases towards the Sun. Since in the present study we are interested only in the agreement of the models at Ulysses location ($\sim 5$~AU from the Sun), where the results of the MC and test particle models agreed to 1\%, we feel justified to adopt in the analysis the values for the PUI production rates at Ulysses calculated with the use of the Warsaw test particle model. 

Earlier versions of the model were described by \citet{rucinski_bzowski:95b, bzowski_etal:97} and \citet{bzowski_etal:02}. The present version of the model was described by \citet{tarnopolski_bzowski:07a} and includes the following effects:
\begin{itemize}
	\item the ionization rate, composed of photoionization, charge-exchange and electron-impact ionization rates; the net ionization rate varies with heliolatitude and its radial profile differs only slightly from $1/r^2$;
	\item the radiation pressure can include the dependence of its magnitude on the radial velocity of individual atoms with respect to the Sun; the net intensity is also variable with time, which results in non-Keplerian trajectories of the atoms;
	\item the inflowing neutral gas is split into two populations of thermal atoms with different parameters at the termination shock.
\end{itemize}

We constrained the simulations by aligning the relevant model parameters with available data wherever possible. Only in lack of available data proxies and models were used to infer the necessary parameters. In order to not disrupt the flow of the discussion, details of the radiation pressure and ionization rate models are presented in the Appendix. In the following section, we will introduce the use of the pickup ion data and how the computation has been adapted to them.

\subsection{Pickup ion data and appropriate computation mesh}

For our comparison we use the local production rate of H$^+$ pickup ions as measured by SWICS/Ulysses \citep{gloeckler_etal:92} from 1997.285 until 1998.310, when Ulysses was crossing the ecliptic plane at aphelion of its orbit, going from $\sim 4.95$ to $\sim 5.4$~AU and descending from $20\degr$ to $0\degr$ ecliptic latitude, which corresponds to an interval from $+12\degr$ to $-6\degr$ heliolatitude  \citep[c.f.][]{gloeckler_geiss:01a}. The spectrum that was used to derive the production rate as a function of distance from the Sun, a quantity that is directly related to the pickup ion flux and thus very close to the observable in this measurement, (as presented in Fig.~\ref{obs2}) is an ensemble average of many individual spectra registered during the time interval mentioned. The individual spectra were selected so that the phase space density in the suprathermal tails ($v/v_{\mathrm{sw}} > 2.4$) was minimum. This eliminated contributions from shocks in the solar wind. The solar wind proton peak was corrected for the instrument dead-time effects. The averaged spectrum was fitted by forward modeling using the classical hot model and the theory of pickup ion transport from \citet{vasyliunas_siscoe:76}. The best fit was obtained for the following set of the hot model parameters: $\mu = 0.9$, $\beta = 6.1\times 10^{-7}$~s$^{-1}$, $v_{\rm TS} = 22$~km/s, $T_{\rm TS} = 12000$~K, $n_{\rm TS} = 0.1$~cm$^{-3}$. This procedure returned a spectrum that then was expressed as the PUI production rate normalized to 1~AU from the Sun. The absolute PUI production rates as a function of distance from the Sun that are shown in Fig. \ref{obs2} were obtained by multiplying this fitted spectrum by by $1/r^2$. The PUI production rate used to derive the H density at the TS was taken from the portion $1.92 < v/v_{\mathrm{sw}} < 2.01$ in the spectrum, which corresponds to the distances marked in Fig. \ref{aa}. Its magnitude was determined to be equal to $7.26\times 10^{-10}$~cm$^{-3}$~s$^{-1}$ with an experimental uncertainty of $\pm 25$\%, almost entirely attributed to the systematic uncertainty in the geometric factor of SWICS \citep{gloeckler_etal:08a}, because statistical fluctuations are negligible for the long time average used here.

Fluctuations in the pickup ion fluxes due to transport effects on shorter time scales do not affect the pickup ion production rate based on the full observation interval. In the following we use this value to determine the density of neutral interstellar H at the termination shock in a comparison with the simulated pickup ion production rate, while varying the ionization rate and radiation pressure within the range of recent observations.

In order to adapt the simulations to the long observation interval with changing locations of Ulysses we performed the calculations of the local hydrogen density on a mesh of $N=16$ points distributed evenly in time along the Ulysses orbit during the observation interval. The densities of the two populations were computed separately for each of the two populations at the 16 points and then combined to obtain the net local density for a given moment of time:
\begin{equation}
\label{eb}
n_{i}\left(r\right)= n_{\mathrm{pri},i}\left(r, \lambda_{i}, \phi_{i},t_{i}\right)+n_{\mathrm{sec} ,i}\left(r,\lambda_{i}, \phi_{i},t_{i}\right),
\end{equation}
where $n_{\mathrm{pri},i}, n_{\mathrm{sec},i}$ are local densities of the primary and secondary populations, $r$ is the Ulysses radial distance, $t_{i}$ is the i-th time moment and $\lambda_{i}$, $\phi_{i}$ are Ulysses heliocentric coordinates at $t_{i}$ (see Fig. \ref{obs1}). Further, the production rates calculated at each of the 16 points were averaged. Therefore, the resulting mean production rate for a given simulation was calculated as follows: 
\begin{equation}
\label{eba}
\beta_{\mathrm{prod}} = \sum_{i=1}^{N} \left[n_{i}\left(r,\lambda_{i}, \phi_{i},t_{i} \right) \, \beta\left(r, \phi_{i}, t_{i} \right)\right]/N,
\end{equation}
where $\beta$ is the net local ionization rate of neutral hydrogen.

\subsection{Calculations}

For the boundary conditions in the Local Interstellar Cloud (LIC) we adopted the gas inflow direction, bulk velocity, and gas temperature as derived by the ISSI Working Group on \textit{Neutral Interstellar Helium} \citep{mobius_etal:04a} from in situ observations of neutral interstellar He atoms \citep{witte:04}, from measurements of He pickup ions  \citep{gloeckler_etal:04a}, and from UV observations of the heliospheric He glow \citep{vallerga_etal:04a}. The upwind direction adopted in the simulations was $\lambda_{B}=254.68\degr$, $\phi_{B}=5.31\degr$ in the B1950.0 ecliptic coordinates; bulk velocity $v_{B}=26.4$~km~s$^{-1}$; and temperature $T_{B}=6400$~K. The density of neutral He in the LIC was adopted as equal to 0.015~cm$^{-3}$, and based on the He ionization degree in the LIC, inferred by \citet{wolff_etal:99} on the level of $\sim 30 - 40$\%, the density of He$^+$ in the LIC was taken equal to 0.008~cm$^{-3}$. 

In our simulations we used two parameter sets for LIC H (referred to by 1. and 2. below),  adopted from \citet{izmodenov_etal:03a} and \citet{izmodenov_etal:03b}, which include:
 \begin{enumerate}
  \item LIC proton density  $n_{p}=0.06$~cm$^{-3}$ and neutral gas density $n_{\mathrm{H}}=0.18$~cm$^{-3}$, which yields the H ionization degree in the LIC equal to 25\%; the contribution of He$^+$ to the net plasma density in the LIC would hence be on the level of 12\%. 
\item LIC  proton density $n_{\mathrm{p}}=0.032$~cm$^{-3}$ and neutral gas density $n_{\mathrm{H}}=0.2$~cm$^{-3}$, with the same density of He$^+$, which yields the H ionization degree of only 14\%.
 \end{enumerate}

Running the Moscow MC model as the first step of the simulation resulted in the following parameters of the primary and secondary populations at the nose of the termination shock, which were adopted in the second step of the simulations with the use of the Warsaw test-particle code:
\begin{enumerate}
{\item Primary: $n_{\mathrm{TS, pri}}=0.1925\, n_{\mathrm{H}} = 0.03465$~cm$^{-3}$, $v_{\mathrm{TS, pri}}=1.08\,v_B = 28.512$~km~sec$^{-1}$, $T_{\mathrm{TS, pri}}= 6020$~K}. 

{\noindent Secondary: $n_{\mathrm{TS, sec}}=0.3345\,n_{\mathrm{H}} = 0.06021$~cm$^{-3}$, $v_{\mathrm{TS, sec}}=0.71 v_B = 18.744$~km~sec$^{-1}$, $T_{\mathrm{TS, sec}}= 16300$~K}. 

The resulting net density at the termination shock was thus equal to $n_{\mathrm{H, TS,I}}=0.53\,n_{\mathrm{H}} = 0.095$~cm$^{-3}$.

{\item Primary:  $n_{\mathrm{TS, pri}}=0.2926\,n_{\mathrm{H}}=0.05852$~cm$^{-3}$, $v_{\mathrm{TS, pri}}=1.07 \,v_{B}=28.248$~km/s; $T_{\mathrm{TS, pri}}=6100$~K;}

\noindent{Secondary: $n_{\mathrm{TS, sec}}=0.2934 \,n_{\mathrm{H}}=0.05868$~cm$^{-3}$, $v_{\mathrm{TS, sec}}=0.7 \,v_{B}=18.48$~km/s; $T_{\mathrm{TS, sec}}=16500$~K.}

The resulting net density at the termination shock was thus equal to $n_{\mathrm{H, TS,II}}=0.59\,n_{\mathrm{H}} = 0.117$~cm$^{-3}$ and the contribution of He$^+$ to the plasma density in the LIC would be much higher, namely 20\%.

\end{enumerate}

In the second step of the simulations, the H densities and H$^+$ PUI production rates at Ulysses were calculated with the Warsaw test-particle model using Eqs (\ref{eb}) and (\ref{eba}), with the parameters of the two populations at the TS as boundary conditions. In general, these models can be subdivided into two groups, depending on the treatment of temporal variations in radiation pressure and ionization rate. In one of the groups, the parameters were calculated as monthly averages for the 16 time intervals $t_i$, and the test-particle program was run in the static mode for each time interval $t_i$, with the parameters pertinent to $t_i$ (``instantaneous'' simulation). In the second group, the parameters were taken as ``smooth'' analytic models, as presented in the Appendix, and the test-particle program was run in its time-dependent mode for each time interval $t_i$ (``smooth'' simulation). 

With such an approach we could assess the influence of short-time fluctuations on the averaged result, which turned out to be negligible (though differences between ``smooth'' and ``instantaneous'' values at $t_i$'s were on the order of 30\%). The result obtained in a comparison of monthly averages and a smooth temporal variation on timescales of one year and longer also justifies our simplification to ignore completely time scales shorter than one month. Since the ``instantaneous'' simulation was much less computer-intensive, most of the further tests were run in the ``instantaneous'' mode. 

Additional variable elements in this part of the simulations were the treatment of the radiation pressure (either dependent on or independent of the radial velocity of the atoms), the inclusion or exclusion of a latitudinal anisotropy in the ionization rate, and inclusion or exclusion of the electron impact ionization.  

\noindent{\em Variation of heliospheric control parameters}

In a next step of the simulations, the robustness of the H$^+$ PUI production rate at Ulysses against variations in the absolute values and treatment of the radiation pressure, ionization rate, and in the modeling strategies was tested. We have repeated the second step of the simulations with various ionization rate and radiation pressure values within the limits of their observational uncertainties. 

These tests included reducing the ionization rate compared with the most recent compilation. Since the charge exchange rates as recorded on Ulysses close to the ecliptic (normalized to 1~AU) appear systematically lower than the values obtained from the OMNI~2 time series, we repeated step 2 simulations with the equatorial charge exchange values $\beta_{\mathrm{eqtr}}$ (see Appendix) reduced to Ulysses measurements, which resulted in an overall reduction of the ionization rate by $\sim 25$\%.

\begin{figure}
\resizebox{\hsize}{!}{\includegraphics[width=8cm]{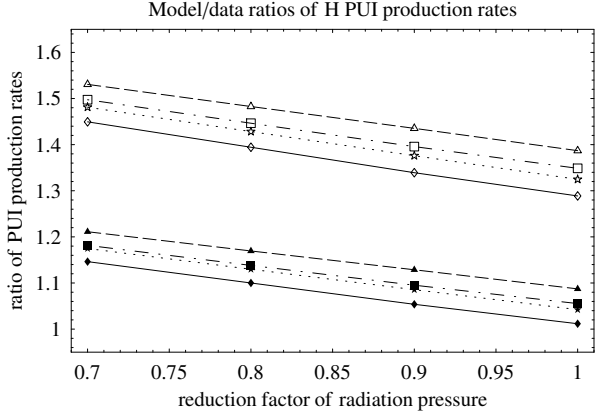}}

\caption{Model/data ratios $S_{i,1}/S_{\mathrm{obs}}$ and $S_{i,2}/S_{\mathrm{obs}}$ of the H$^+$ PUI production rates at Ulysses as a function of the reduction factor of the solar radiation pressure relative to the currently adopted value. The group of lines with filled symbols (1) corresponds to the simulations performed with the density at the termination shock equal to 0.095~cm$^{-3}$, the open symbols (2) correspond to 0.117~cm$^{-3}$. Diamonds correspond to the nominal values of the ionization rate and  radiation pressure dependent on $v_{r}$, stars to the reduced values of the ionization rate and radiation pressure dependent on $v_{r}$, and squares and triangles to radiation pressure independent on $v_{r}$, with the ionization rate, respectively, nominal and reduced. The observed PUI production rate was equal to  $7.26\times10^{-10}$~cm$^{-3}$~s$^{-1}$.}
\label{ra}
\end{figure}

Since the absolute calibration of the solar Lyman-$\alpha$ input has changed appreciably since the beginning of the observations in the 1970-ties, when it was believed that the effective radiation pressure factor at solar minimum was $\mu \simeq 0.7$ and at solar maximum $\mu \simeq 1$ \citep[e.g.][]{vidal-madjar:75, tobiska_etal:97}, compared to present-day values $\simeq 1$ at solar minimum and $\simeq 1.5$ at solar maximum  \citep{woods_etal:00, tobiska_etal:00c}, we repeated step-2 calculations for the line-integrated flux reduced by factors 0.9, 0.8, and 0.7, as illustrated in Fig.\ref{ra}. We also checked for robustness against uncertainties in the solar Lyman-$\alpha$ line profile by performing simulations with a simplified flat line-center (hence, the radiation pressure independent of atom radial radial velocities) and a self-reversed profile (with radiation pressure dependent on the radial velocity).

As can be seen from the previous discussion, the variations in the ionization rate and  radiation pressure introduced into our simulations were not chosen arbitrarily, and they did not exceed uncertainties related to changes in calibrations and increased measurement sophistication. 

To also assess the sensitivity of the results to uncertainties in the cross section for charge exchange, we repeated the simulations for the first LIC parameter set with the \citet{maher_tinsley:77} cross section replaced with that from  \citet{lindsay_stebbings:05a}. As discussed by \citet{fahr_etal:07a}, the two formulae agree to a few percent in the supersonic solar wind regime, but differ up to $\sim 40$\% for low collision energies, pertinent to the region between the heliopause and the bow shock, where the primary interstellar population loses a portion of its atoms and the secondary population is created due to charge exchange with protons. As illustrated by \citet{baranov_etal:98a}, the change in coupling between the protons and H atoms results in different proportions between the primary and secondary populations at the termination shock. Our simulations showed that apart from the changes in the individual densities of the populations at the termination shock, their remaining parameters, i.e. bulk velocities and temperatures, change very little:

{\noindent Primary: $n_{\mathrm{TS, pri}}=0.02592$~cm$^{-3}$, $v_{\mathrm{TS, pri}}=28.776$~km~sec$^{-1}$, $T_{\mathrm{TS, pri}}= 5900$~K,} 

{\noindent Secondary: $n_{\mathrm{TS, sec}}=0.07020$~cm$^{-3}$, $v_{\mathrm{TS, sec}}=18.520$~km~sec$^{-1}$, $T_{\mathrm{TS, sec}}= 16500$~K;\\ 
the net density at the termination shock was equal to 0.096~cm$^{-3}$, i.e. practically identical as the in the simulation with the old cross section formula.}

Hence, the propagation and losses of the two populations during their travel from the termination shock to the inner heliosphere were almost identical in the cases of Maher \& Tinsley and Lindsay \& Stebbings formulae, but the input values to the simulation inside the termination shock were different.

\section{Results}
The results of the extended simulations are compiled in Fig.\ref{ra} as ratios of the simulated production rates to the measured value. The simulations show that indeed, as postulated in Section 2, the model H$^+$ PUI production rate at the location of Ulysses during the observation interval is only weakly dependent on the radiation pressure, ionization rate, and details of modeling of the gas density in the inner heliosphere. A change in the ionization rate by $\sim 25$\% (between diamonds and stars or triangles and squares in Fig.\ref{ra}) results in a change in the density at the termination shock of only $\sim 2.5$\%. The combined variation in the H$^+$ PUI production rate due to details of the ionization rate and the modeling approach for radiation pressure does not amount to more than 4\%. Varying the level of solar Lyman-$\alpha$ output by 30\% translates into a somewhat larger variation in the production rate, on the level of 10\%, but again this is substantially weaker than the variation in the input. Overall, the amplitude of the variations in each of the two input factors is reduced by a factor of 10 for the ionization rate and 3 for the radiation pressure in the resulting variations in the inferred interstellar H density at the termination shock.

This result is also robust against other details of the simulations, including the solar Lyman-$\alpha$ line shape (inverted profile vs flat), functional form of the charge exchange cross section (\citet{maher_tinsley:77} vs \citet{lindsay_stebbings:05a}),  treatment of temporal variations in radiation pressure and the ionization rate (fully time-dependent vs static with solar-rotation averaging), presence or absence of the latitudinal variations in the ionization rate, and presence or absence of the electron ionization. All these factors affect the production rate at Ulysses during the observation period only by a few percent and are not able to significantly change its value. 

To calculate the density at the termination shock, we take the PUI production rates $S_{i,1}$ for the nominal $\mu$ values and the $\mu$ values reduced by 10\%, obtained from the first simulation (filled symbols in Fig. \ref{ra}), and we we calculate the density at the termination shock $n_{\mathrm{H,TS}}$ from the formula:
\begin{equation}
\label{eca}
n_{\mathrm{H,TS}} = \frac{n_{\mathrm{H,TS},1}}{\left<S_{i,1}/S_{\mathrm{obs}}\right>},
\end{equation}
where the angular brackets denote arithmetic mean, $S_{\mathrm{obs}}$ is the PUI production rate from the Ulysses observations discussed in this paper, $S_{i,1}$ are the production rates from the first (1) set of simulations, and $n_{\mathrm{H,TS},1}$ is the density at TS assumed in the first simulation (denoted with index 1). The $n_{\mathrm{H,TS}}$ value obtained from this calculation is equal to $0.089\pm0.003$~cm$^{-3}$. 

The robustness of this result is supported by the derivation of the density at the TS from the other simulation (2), for which the TS density for the nominal set of ionization and radiation pressure values was assumed to be 0.117~cm$^{-3}$ (the results are shown as open symbols in Fig.~\ref{ra}). Deriving the TS density analogous to the description in the previous paragraph yields $n_{\mathrm{H,TS}} = 0.086 \pm 0.003$~ cm$^{-3}$, i.e. very close to the previous value although the starting TS density in the simulation was higher by 35\%. 

Hence, by taking the average of the two values, we arrive at the density of interstellar hydrogen at the TS of 0.087~cm$^{-3} \pm 25$\%, where the uncertainty is almost completely dominated by the instrumental uncertainty of the absolute geometric factor.

\section{Discussion and conclusions}

We have used the accumulation of the H$^+$ pickup ion production rate from SWICS/Ulysses over a $\sim 13$~month period in 1997 -- 1998 at the Ulysses passage through the solar equator plane to infer the interstellar H density at the termination shock. By extensive simulations we demonstrated that the H$^+$ PUI production rate in this location of the heliosphere is only weakly dependent on the values of solar radiation pressure and neutral H ionization rate, but sensitively depend on the density at the termination shock. We have found that the H density at the termination shock inferred from these pickup ion production rates is very robust against any variations in the ionization rate, radiation pressure, and the actual modeling approaches for the density distribution in the inner heliosphere. 

In the present analysis we have, for the first time, included explicitly both the observational uncertainty and the modeling uncertainties. While in our approach the modeling uncertainties are minimized to a few percent, a larger uncertainty is incurred for the observation because absolute flux values are used, which results in an uncertainty of the obtained termination shock density equal to $\pm 25$\%. In the previous approach by \citet{gloeckler_geiss:01b} the observational uncertainty was minimized by making use of the ratio of the pickup ion and solar wind flux, but the $\sim 10$\% uncertainty quoted in \citet{izmodenov_etal:03a} does not include a range of values for the ionization rate and radiation pressure and the uncertainty of the geometric factor of the instrument. But, as demonstrated in the previous sections, the density at the termination shock scales linearly with the observed pickup ion production rate, which is directly related to the pickup ion flux and/or distribution function. Hence, any observational uncertainties will transfer linearly into the resulting densities. Since \citet{izmodenov_etal:03a} started from the local neutral gas density at Ulysses, any uncertainty in the ionization rate will appear approximately linearly in the extrapolated density at the termination shock. 

The determination of the PUI production rate at Ulysses near the aphelion, on which our derivation of $n_{\mathrm{H,TS}}$ is based, is not entirely model-free. Although in the present determination of the PUI production rate at Ulysses a simple hot model was used for forward-modeling of the PUI distribution function, our method is robust against simplifications inherent to that kind of modeling because it uses a quantity (i.e., the production rate of PUI at Ulysses) which weakly depends on details of such modeling.

The determination of the H density at the termination shock by \citet{gloeckler_etal:08a}, equal to $0.080\pm0.008$~cm$^{-3}$, is free of the uncertainty of the geometric factor of the instrument, but is subject to a combination of uncertainties in the determination of the He density and that of the He abundance relative to H from the Voyager LECP observations, including uncertainties in the ratios of the production rates of these species. Nevertheless, after including of all uncertainties, all three approaches (i.e. the present one and those from \citet{gloeckler_geiss:01b} and \citet{gloeckler_etal:08a}) should be read with a similar uncertainty band. The density value presented here agrees very well with the new determination by \citet{gloeckler_etal:08a}. Although our value still agrees with the previous determination of the H density from SWICS pickup ion observations by \citet{gloeckler_geiss:01b} within their mutual uncertainty bands, the combination of the two new results suggest a somewhat lower density than 0.1~cm$^{-3}$. 

Our results also agree comfortably with the TS density values found from the analysis of the heliospheric Lyman-$\alpha$ glow \citep[this volume]{pryor_etal:08a} and from the solar wind slowdown \citep[this volume]{richardson_etal:08a} within the uncertainty bands. It should be noted here that the coupling between the neutral and ionized component of the interstellar medium between the bow shock and the heliopause appears to be somewhat stronger than suggested previously. This is a consequence of an updated relation for the energy dependence of the charge exchange cross-section between protons and H atoms \citep{lindsay_stebbings:05a}. 

The parameters of the interstellar gas in front of the heliospheric bow shock, as assumed in simulation (1), also seem to be robust, as shown by \citet[this volume]{mueller_etal:08a}, who discussed the present status of the modeling of heliospheric interface and showed that differences in the filtration rate returned by different models of the heliosphere evaluated with identical initial parameters are about 15\% and this result can be adopted as the uncertainty of the H density in the CHISM. 

\begin{appendix}
\section{Inner-heliospheric environment during observations }

\subsection{Radiation pressure}
\begin{figure}
\centering
\includegraphics[width=8cm]{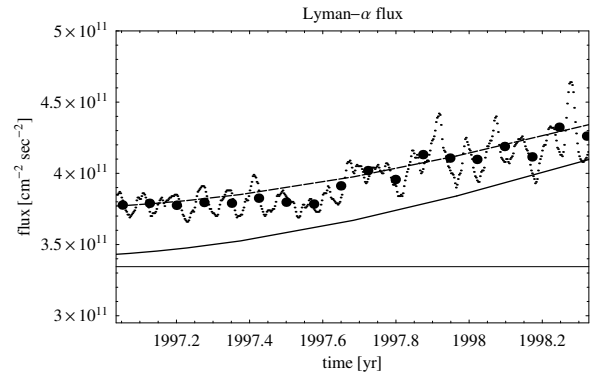}
\caption{Daily values of line-integrated solar Lyman-$\alpha$ flux (small dots) from SOLAR 2000, Carrington-averaged values (thick dots) and fitted models (smooth lines). The upper (broken) line corresponds to the approximation of the Carrington-averaged time series by Eq.(\ref{eb1}). The lower (solid) line indicates the effective radiation pressure after rescaling to the central band with Eq.(\ref{ea1}). The horizontal line corresponds to $\mu = 1$.}
\label{xmu}
\end{figure}

Radiation pressure varies with heliocentric distance, radial velocity of individual atoms, and time. Both of the cases which will be discussed below are based on the line- and disk-integrated solar Lyman-$\alpha$ flux $I_{\mathrm{tot}}$ at 1~AU (expressed in photons~cm$^{-2}$~s$^{-1}$), based on daily values obtained from the SOLAR 2000 model \citep{tobiska_etal:00c}, which are shown for the observation interval as small dots in Fig. \ref{xmu}. 

Because it takes years for the interstellar flow to pass even only through the inner heliosphere, we ignored variations on time scales shorter than one solar rotation period (i.e., a month), which is further justified by the fact that two different treatments on longer time scales lead to almost identical results. To assess the influence of slower temporal variations on the modeling we constructed two models: ``instantaneous'' and  ``smooth''. In the ``instantaneous'' model the appropriate monthly values, shown as thick dots in Fig. \ref{xmu}, were taken within a time-independent model for the entire month. In the ``smooth'' model, the variation in $I_{\mathrm{tot}}$ was approximated, following \citet{bzowski:01a}, by the relation:
\begin{equation}
\label{eb1}
I_{\mathrm{tot}}\left (t\right)=I_{\mathrm{tot},0}+\sum _{i=1}^{N_{\mu }} \left(a_i\cos  \omega _it + b_i\sin  \omega _it\right)
\end{equation}
with $N_\mu =7$. The relevant frequencies and their amplitudes were obtained from the analysis of the Lomb periodogram \citep{press_rybicki:89} of a time series composed of the Carrington-averages of $I_{\mathrm{tot}}$ for the time period 1948--2004 (results obtained using daily values are almost identical). Periodicities shorter than one year and amplitudes smaller than 0.025 of the strongest harmonic were ignored. The parameters obtained are listed in Table~\ref{muTable} and the resulting curve is shown as the upper broken line in Fig. \ref{xmu}.

\begin{figure}
\resizebox{\hsize}{!}{\includegraphics[width=8cm]{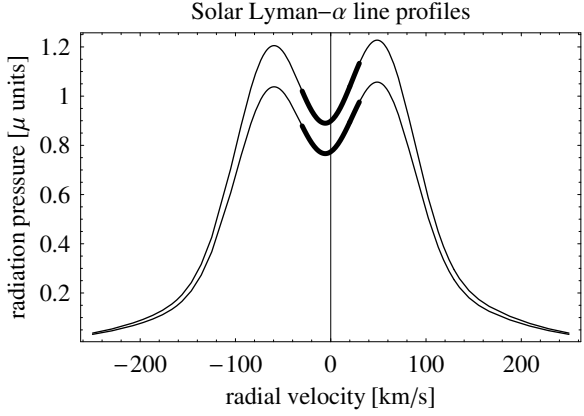}}
\caption{Model solar Lyman-$\alpha$ line profiles, based on data from \citet{lemaire_etal:02}, for the start and end of the PUI observation interval, expressed in the $\mu$ units. Thick lines indicate the approximate radial velocity range for the H atoms at Ulysses, adopted as the bulk velocity $\pm 3 \times$~the thermal speed. The range is used for the ``flat'' model in Eq.(\ref{ea1}).}
\label{x1}
\end{figure}

The solar Lyman-$\alpha$ line has a two-peaked, self-reversed profile (Fig. \ref{x1}). The radiation pressure acting on individual H atoms depends on the Doppler shift resulting from their radial motion relative to the Sun. Therefore, the Lyman-$\alpha$ spectral flux was converted into the radiation pressure factor $\mu$, using either a ``flat'' or a ``Doppler'' model. 

\begin{figure}
\centering

\includegraphics[width=6cm]{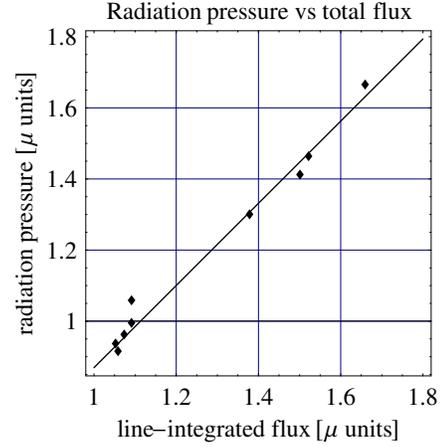}
\caption{Relation between the line-integrated flux of the solar Lyman-$\alpha$ output, expressed in the $\mu$-units, and the coefficient $\mu$ of radiation pressure acting on interstellar hydrogen atoms in the inner heliosphere. Dots correspond to the values obtained from averaging of the 9 solar Lyman-$\alpha$ profiles observed by \citet{lemaire_etal:02} over $\pm 30$~km/s about the line center. The line is a fit defined in Eq.(\ref{ea1}).}
\label{xg}
\end{figure}

Since most of the H atoms at $\sim 5$~AU do not exceed $\sim 30$~km/s, one can approximate the spectral flux responsible for the radiation pressure by averaging the line profile over $\sim \pm 30$~km/s about the line center for a few solar line profile data sets and relating these values to the routinely-measured $I_{\mathrm{tot}}$. Based on observations by \citet{lemaire_etal:02}, the averaged central Lyman-$\alpha$ flux correlates linearly with the total flux $I_{\mathrm{tot}}$, as illustrated in Fig. \ref{xg}, and a relation for the ``flat'' radiation pressure factor $\mu_{\mathrm{flat}}$ can be given as:
\begin{equation}
\label{ea1}
\mu _{\mathrm{flat}}= 3.473\times 10^{-12}\, I_{\mathrm{tot}}-0.287.
\end{equation}
A similar fit was published by \citet{emerich_etal:05} for the center of the line profile, i.e. radial velocity of 0 km/s. The ``flat'' approximation for the radiation pressure has been used in connection both with the ``instantaneous'' and ``smooth'' time series of the Lyman-$\alpha$ fluxes.

The ``Doppler'' model of the radiation pressure is based on a functional fit to the 9 solar Lyman-$\alpha $ line profiles observed by \citet{lemaire_etal:02}. The model, discussed in detail by \citet{tarnopolski_bzowski:07a}, uses the functional form:
\begin{eqnarray}
\label{e3}
\mu \left(v_r, I_{\mathrm{tot}}\right)&=&A\left(1 + B\, I_{\mathrm{tot}}\right) \exp \left(-C v_r^2\right)\\ &\times &\left[1 + D \exp \left(F v_r
- G v_r^2\right) + H \exp \left(-P v_r - Q v_r^2\right)\right] \nonumber
\end{eqnarray}
where $v_{r}$ is the radial velocity of a H atom in km/s. The parameters of the fit are compiled in Table \ref{profTable}. Sample fits for the beginning and the end of the pickup ion observation interval are shown in Fig. \ref{x1}. 

In summary, four baseline models were compared, i.e. the ``flat'' approximation and the ``Doppler'' model of the radiation pressure in combination with the ``instantaneous'' and ``smooth'' temporal dependences. The absolute variation in the radiation pressure discussed in the main text was introduced by multiplying each model by scaling factors ranging from 0.7 to 1.

With these in hand, one can construct the radiation pressure for a given time $t$ and radial velocity $v_r$ by inserting of the $I_{\mathrm{tot}}\left(t\right)$ value to Eq.(\ref{e3}), obtained from either the smooth or instantaneous model. 
 
\begin{table}
\caption{Parameters of Eq.(\ref{xmu}) needed to calculate the line-integrated solar Lyman-$\alpha$ flux for time $t$ in years; $I_{\mathrm{tot}, 0} = 4.6034\times 10^{11}$~cm$^{-2}$~s$^{-1}$.}
\label{muTable}
\centering
\begin{tabular}{c c c c}
\hline \hline
$i$ & $\omega_{i}$ & $a_{i}$ & $b_{i}$ \\
\hline
 1  & $0.14406$ & $-1.7673\times 10^{10}$ & $-1.5657\times 10^{10}$ \\
 2  & $0.31182$ & $-6.3068\times 10^{9}$  & $ 1.0732\times 10^{10}$ \\
 3  & $0.43745$ & $-2.1161\times 10^{9}$  & $ 8.3570\times 10^{9}$ \\
 4  & $0.58427$ & $ 5.0496\times 10^{10}$ & $ 7.9792\times 10^{10}$ \\
 5  & $0.74027$ & $-2.0954\times 10^{10}$ & $-9.6306\times 10^{9}$ \\
 6  & $1.14430$ & $-1.8691\times 10^{10}$ & $-1.3267\times 10^{9}$ \\
 7  & $1.96351$ & $-6.2468\times 10^{9}$  & $-3.8329\times 10^{9}$ \\
\hline
\end{tabular}
\end{table}
\begin{table}
\caption{Parameters of the model of radiation pressure dependence on radial velocity $v_r$ and total flux $I_{\mathrm{tot}}$ defined in Eq.(\ref{e3}).}
\label{profTable}
\centering
\begin{tabular}{lllllllll}
\hline
 $A  =  2.4543 \times 10^{-9}$, & $B =  4.5694\times 10^{-4}$, & $C =  3.8312\times 10^{-5}$, \\ $D =  0.73879$, & $F =  4.0396\times 10^{-2}$, &  $G =  3.5135\times 10^{-4}$, \\ 
 $H =  0.47817$, &  $P =  4.6841\times 10^{-2}$, &  $Q =  3.3373\times 10^{-4}$\\
\hline
\end{tabular}
\end{table}

Hence, a total of 4 baseline models of radiation pressure were exercised: two ``flat'' models, with $I_{\mathrm{tot}}$ either instantaneous or smooth, and two ``Doppler'' models, also with $I_{\mathrm{tot}}$ either instantaneous or smooth. The reduction of radiation pressure discussed in the main text was executed by multiplying relevant model by scaling factors ranging from 0.9 to 0.7.

\subsection{Ionization processes}

Ionization of H in interplanetary space occurs through a combination of three processes. Charge exchange with solar wind and photoionization by solar EUV are usually two major contributors and both scale almost perfectly with the inverse square of the distance $r$ from the Sun. Electron impact ionization is mostly a small contribution that increases closer to the Sun (inside $\sim 3$~AU), but becomes completely negligible at large distances.

\subsubsection{Charge exchange}

The charge exchange rate within the supersonic solar wind was calculated in the standard way \citep[e.g.][]{bzowski:01b}, with the use of the widely adopted formula:
\begin{equation}
\label{ecx}
\beta_{\mathrm{cx}}=\sigma_{\mathrm{cx}}\left(v_{\mathrm{SW}}\right) n_{p} v_{\mathrm{SW}}
\end{equation}
where $v_{\rm{SW}}$ is the solar wind speed, $n_{\text{SW}}$ solar wind density, and $\sigma _{\text{cx}}$ is the reaction cross section, which depends on the relative velocity of the particles. At a given heliolatitude the solar wind speed was assumed to be constant up to the termination shock, with the density to drop off as $1/r^2$. Hence the charge exchange rate decreases with $1/r^2$.  

Similarly to our treatment of radiation pressure, two approaches were used to model the time variations: ``smooth'' and ``instantaneous''. As input we used the equatorial rate at 1~AU, denoted $\beta_{\mathrm{eqtr}}$. Another variable in the model was the presence or absence of latitudinal anisotropy. This resulted in 4 baseline models: spherically symmetric ``smooth'' or ``instantaneous'' and latitude-dependent ``smooth'' or ``instantaneous''. In the latitude-dependent model the anisotropic part of the model was calculated as shown below, for any given time $t$. In the ``instantaneous'' case the latitudinal structure was ``frozen'' as for this time, while in the ``smooth'' case a fully time-dependent model was realized, with the latitudinal rate evolving along the trajectory of the atoms. 

{\noindent \em Evolution of the equatorial rate}

\begin{figure}
\centering
\includegraphics[width=8cm]{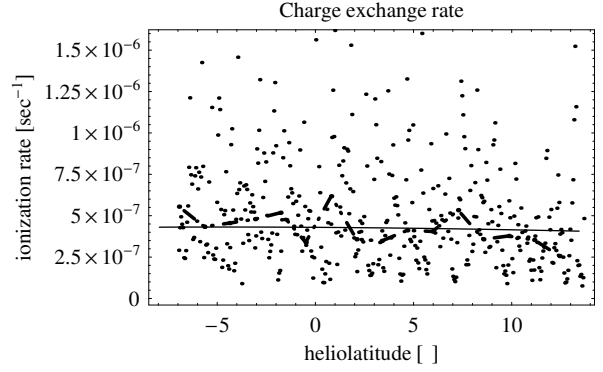}

\caption{Comparison of the model equatorial charge exchange rates with the Ulysses observations obtained during the PUI measurementsinterval. The broken line is the ``instantaneous'' model,  the solid line is the ``smooth'' model (Eq. (\ref{eb1})), and dots are the OMNI-2 daily rates.}
\label{xs}
\end{figure}

The approach to model the equatorial rate of charge exchange was similar to the modeling of the line-integrated flux of the solar Lyman-$\alpha$ radiation in Section A1, now based on the daily values of the solar wind speed and density from the OMNI-2 collection \citep{king_papitashvili:05}, normalized to 1~AU. The charge exchange rate was calculated according to Eq.(\ref{ecx}) and subsequently averaged over Carrington periods. The results were taken as the ``instantaneous'' model for the relevant 16 time intervals. For the ``smooth'' model a periodogram analysis was performed, which returned a formula similar to Eq.(\ref{eb1}), with $N_{cx} = 5$ periodicities and the remaining parameters collected in Table~\ref{tabBetEqtr}. The results are shown in Fig. \ref{xs}. 

A comparison of the daily charge exchange rates based on OMNI-2 with observations on Ulysses (scaled to 1~AU) showed systematic differences, with the Ulysses values lower by $\sim 25$\%. This resulted in another set of models, with the rates reduced by 25\% relative to the OMNI-2 values.  

\noindent{\em General formula for the 3D charge exchange rate}

To perform a realistic simulation of the Ulysses PUI observations one has to include the  evolution in the charge exchange rate with heliolatitude. Based on observations, \citet{bzowski:01b} proposed an analytical phenomenological formula for the charge exchange rate at 1~AU, repeated here for reader's convenience in a slightly modified form:
\begin{eqnarray}
\label{ec}
\beta_{\mathrm{cx}} \left(\phi ,t\right)&=&\left(\beta_{\mathrm{pol}} + \delta_{ \beta}\, \phi \right) + \left(\beta _{\mathrm{eqtr}}\left(t\right)-\beta_{\mathrm{pol}}\right)\\  & \times &\exp \left[-\ln  2 \left(\frac{2 \phi  - \phi_{N}\left(t\right)- \phi_{S}\left(t\right)}{\phi_{N}\left(t\right)-\phi_{S}\left(t\right)}\right)^n\right].\nonumber
\end{eqnarray}
$\phi $ is the heliographic latitude, $n$ a shape factor which was adopted as $n=2$, $\beta_{\mathrm{pol}}$ the mean charge exchange rate at the poles. The term $\left(\beta_{\mathrm{pol}}+ \delta_{ \beta}\, \phi \right)$ describes the north-south asymmetry of the polar rates, and the term $\exp\left[-\ln 2 \left(\frac{2 \phi-\phi_{N}-\phi_{S}}{\phi_{N}-\phi_{S}}\right)^n\right]$ provides the latitudinal evolution. 

{\noindent \em Evolution of latitudinal anisotropy}

Ulysses in situ observations \citep{mccomas_etal:99} were not sufficient to infer the evolution in the latitudinal anisotropy and had to be supplemented by observations of the heliospheric Lyman-$\alpha$ glow, which is sensitive to the structure of the solar wind. As demonstrated by SWAN/SOHO \citep{bertaux_etal:99}, the glow features a darkening in an ecliptic band during low solar activity, nicknamed the heliospheric groove. The groove is due to the latitudinal anisotropy of the solar wind, making it a good tracer for the latitudinal structure \citep{bzowski:03}. \citet{bzowski_etal:03} exploited this to infer the equatorial-to-pole contrast of the charge exchange rates and the latitudinal boundaries of the polar regions of reduced rates for selected dates between solar minimum and maximum. 

With this information and continuous coverage of the equatorial charge exchange rate, ``snapshot pictures'' of the charge exchange ionization field in the inner heliosphere were worked out.

Gaps in the coverage of the Lyman-$\alpha$ images were filled by observations of polar holes reported by \citet{harvey_recely:02}, who provide a time series of the latitudinal boundaries of polar holes between two consecutive solar maxima. 

\begin{figure}
\centering
\includegraphics[width=8cm]{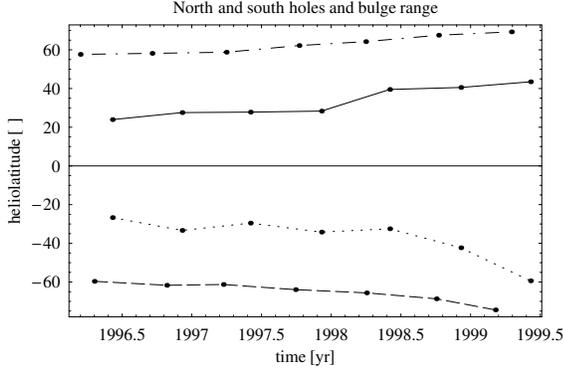}
\caption{Latitudinal boundaries of coronal holes as reported by \citet{harvey_recely:02} (top and bottom lines) and boundaries of the ``equatorial bulge'' of the charge exchange rate from \citet{bzowski_etal:03} (solid and dotted lines).}
\label{xm}
\end{figure}

We took advantage of a new linear correlation between the areas of the polar holes from  \citet{harvey_recely:02} $S_{\mathrm{ch,}N}$, $S_{\mathrm{ch,}S}$, defined as follows:
\begin{eqnarray}
\label{ed}
S_{\mathrm{ch,}N}&=&\int _{\phi_{\mathrm{ch,}N}}^{\pi /2}\int _0^{2 \pi }\cos (\phi )\,\mathrm{d}\phi \,\mathrm{ d}\lambda  =2 \pi \left(1 - \sin \left(\phi_{\mathrm{ch,}N}\right)\right) \nonumber \\
S_{\mathrm{ch,}S}&=&\int _{-\pi /2}^{\phi_{\mathrm{ch,}S}}\int _0^{2 \pi }\cos (\phi )\,\mathrm{d}\phi \,\mathrm{d}\lambda =2 \pi \left(1 + \sin \left(\phi_{\mathrm{ch,}S}\right)\right),
\end{eqnarray}
and the areas of reduced charge exchange rate $S_{N}$, $S_{S}$, inferred by \citet{bzowski_etal:03} from observations of the heliospheric glow. $\phi_{\mathrm{ ch,}N}$, $\phi_{\mathrm{ ch,}S}$ are the latitudes of the north and south hole boundaries. Both boundaries are shown in Fig.\ref{xm}.  
Since, as inferred from the observations, these areas vanish at solar maximum, coefficients were fitted separately to the northern and southern areas:
\begin{eqnarray}
\label{ed1}
S_{N} = a_{N}\, S_{\mathrm{ ch,}N} \nonumber \\
S_{S} = a_{S}\, S_{\mathrm{ ch,}S}
\end{eqnarray}
with $a_{N} = 3.91$ and $a_{S} = 4.20$. 

$S_{N}$, $S_{S}$ are computed from Eqs.(\ref{ed1}) using the observed polar hole areas based on the boundaries $\phi_{\mathrm{ch,}N}$, $\phi_{\mathrm{ ch,}S}$. Finally, the boundaries of the reduced ionization rate regions in the northern and southern hemispheres emerge from Eq.(\ref{eg}) as follows:  
\begin{eqnarray}
\label{eg}
\phi_{N}&=&\arcsin \left(1 - 2\, S_{N}\right) \nonumber \\
\phi_{S}&=&-\arcsin \left(1 - 2\, S_{S}\right)
\end{eqnarray}

\begin{figure}
\centering
\includegraphics[width=8cm]{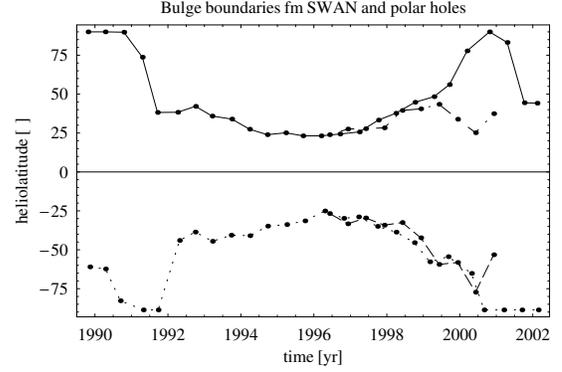}
\caption{Boundaries of the ``equatorial bulge'' in the charge exchange rate based on coronal hole boundaries and Eq.(\ref{eg}) (solid and dotted lines), compared with the boundaries inferred by \citet{bzowski_etal:03} from the heliospheric Lyman-$\alpha $ glow (broken lines).}
\label{xo}
\end{figure}

As shown in Fig.\ref{xo}, this result compares well with the boundaries inferred from SWAN observations. To simplify computations, the boundaries of the coronal holes were further approximated by:
\begin{equation}
\label{efi}
\phi_{N,S}\left(t\right) = \phi_{0} + \phi_{1}\, \exp\left[-\cos^3\left(\omega_{\phi}\, t \right) \right]
\end{equation}
with the parameters listed in Table~\ref{phiTable}.
\begin{table}
\caption{Parameters to calculate the equatorial charge exchange rate for  time $t$ in years; $\beta_{\mathrm{eqtr}, 0} = 5.2978\times 10^{-7}$~s$^{-1}$.}
\label{tabBetEqtr}
\centering
\begin{tabular}{c c c c}
\hline \hline
$i$ & $\omega_{i}$ & $a_{i}$ & $b_{i}$ \\
\hline
 1  & $0.15671$ & $-6.4582\times 10^{-8}$ & $-4.0790\times 10^{-8}$ \\
 2  & $0.39108$ & $ 3.3963\times 10^{-8}$ & $ 3.3016\times 10^{-9}$ \\
 3  & $0.59997$ & $-3.8191\times 10^{-8}$ & $-1.7056\times 10^{-8}$ \\
 4  & $0.96353$ & $-2.5358\times 10^{-8}$ & $ 3.9630\times 10^{-9}$ \\
 5  & $1.21876$ & $-3.6585\times 10^{-8}$ & $-1.9796\times 10^{-8}$ \\
\hline
\end{tabular}
\end{table}
\begin{table}
\caption{Parameters of Eq.(\ref{efi}) needed to calculate the boundaries of polar holes for time $t$ expressed in decimal years}
\label{phiTable}
\centering
\begin{tabular}{c c c c}
\hline \hline
$N/S$ & $\omega_{\phi}$ & $\phi_{0}$ & $\phi_{1}$ \\
\hline
 N  & $0.58251$ & $ 31.2$ & $-24.0$ \\
 S  & $0.58226$ & $-38.7$ & $ 21.5$ \\
\hline
\end{tabular}
\end{table}

{\noindent \em North -- south asymmetry of the charge exchange rate}

The north-south asymmetry in the polar charge exchange rates was discovered during the first Fast Latitude Scan by Ulysses in 1995 \citep{mccomas_etal:99,mccomas_etal:00b, bzowski_etal:03}. Its long-term evolution is still unknown. As an ad hoc solution, we include this effect with the term $\beta_{\mathrm{pol}} + \delta_{ \beta}\, \phi$ in Eq.(\ref{ec}), where $\beta_{\mathrm{pol}} = 2.5405\times 10^{-7}$ and  $\delta_{ \beta} =  -1.7863\times 10^{-10}$. 

\subsubsection{Photoionization}

Over time scales longer than one solar rotation and at distances outside a few dozen solar radii, the photoionization field can be treated as spherically symmetric, with the intensity falling off with the square of the distance. Departures from spherical symmetry are about $\sim 10$\% \citep{auchere_etal:05a} and were neglected. 

\begin{figure}
\centering
\includegraphics[width=8cm]{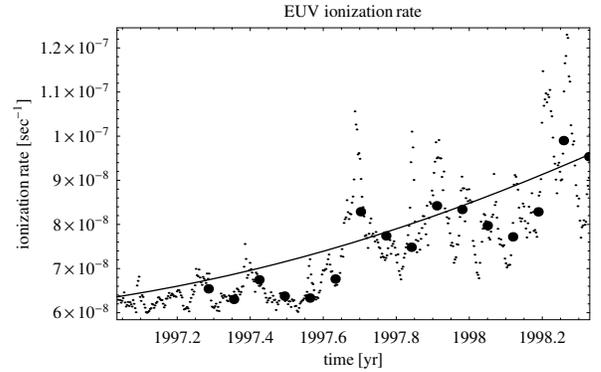}
\caption{EUV ionization rates: ``instantaneous'' (thick dots) and ``smooth'' model (continuous line). Small dots represent daily values as computed from the 10.7~cm proxy. }
\label{xk}
\end{figure}

The photoionization rate $\beta_{\mathrm{ph}}$ was treated similarly to the net Lyman-$\alpha$ flux $I_{\mathrm{tot}}$, with the ``instantaneous'' and ``smooth'' models. Measurements of the photoionization rates of H over the solar cycle are not readily available, so proxies have to be used. A reasonable proxy for the H photoionization rate is the solar radio flux in the 10.7~cm band \citep{bzowski:01a}. Thus daily values of the Ottawa solar 10.7~cm flux were used, converted to daily photoionization rates \citep{bzowski:01a}, and then Carrington-averaged, as shown for the ``instantaneous'' and ``smooth'' modeling in Fig. \ref{xk}. The rate relevant for the Ulysses PUI observations was about $0.8\times 10^{-7}$~s$^{-1}$, slowly increasing with time.  

When this research was already pretty much advanced, a new version of SOLAR 2000 was introduced, with the H photoionization rates deduced from the proxies-inferred solar spectra. A comparison of the rates derived in the two ways showed that a difference in the amplitudes exists, while the solar cycle-averages agree. The ratio of amplitudes is $\sim 1.4$. Using the two predictions for the simulations of the Ulysses observations, we find an agreement of the results within $\sim 10$\%. Because the simulations are time consuming and the influence of photoionization is weak, we decided to keep the model based on the 10.7~cm proxy. Since the proxies used in the SOLAR 2000 model are much more elaborate than the simple 10.7~cm proxy, the H photoionization rate based on SOLAR 2000 is probably an improvement and thus should be used in future studies, in particular during solar minimum and maximum, when the differences are largest.

\subsubsection{Ionization by electron impact}

Throughout the simulation we used only one model of the electron-impact ionization rate. The electron impact rate was calculated based on the radial evolution in the solar wind electron temperature as derived by \citet{marsch_etal:89} from Helios data and on the mean solar wind density taken from \citet{kohnlein:96}. The electron density was adopted assuming quasi-neutrality of the solar wind, as a sum of the proton density and twice the alpha density. 

The electron distribution function can be approximated as a bi-Maxwellian, with a warm core and hot halo populations, plus an occasional strahl population along the local magnetic field line \citep{pilipp_etal:87c, pilipp_etal:87b}. The contribution of the halo population to the net rate is on the level of a few percent and is an increasing function of the heliocentric distance \citep{maksimovic_etal:05a}. The calculations that we performed based on the ACE electron data presented by \citet{mcmullin_etal:04a} show that at 1~AU the ionization rate due to the core population of the solar wind electrons is equal to about $0.4\times 10^{-7}$~s$^{-1}$ and to the halo population to less than $0.04\times 10^{-7}$~s$^{-1}$. Our assessment showed also that the amplitude of fluctuations in the electron ionization rate may reach an order of magnitude, which is much more than the long-time variations related to variations over the solar cycle. On the other hand, the electron data from Wind \citep{salem_etal:03a} lead to an in-ecliptic solar minimum (1995) rate of $\sim 0.68\times 10^{-7}$~s$^{-1}$ and a solar maximum (2000) rate of $\sim 0.73\times 10^{-7}$~s$^{-1}$. Thus assuming a constant rate over the solar cycle is a good approximation.

\begin{figure}
\centering
\includegraphics[width=8cm]{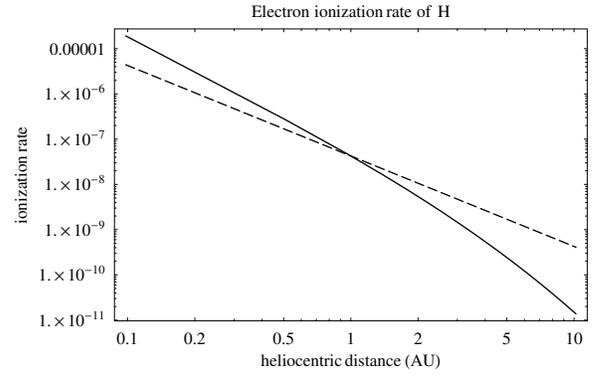}

\caption{Radial profile of the electron-impact ionization rate of neutral interstellar H used in the simulations (solid line). The radial dependence of this rate differs significantly from the $1/r^{2}$ law valid for charge exchange and photoionization, as illustrated by the broken line.}
\label{xj}
\end{figure}

Because of the lack of electron distribution data during the observation interval, the rate was calculated assuming a mono-Maxwellian electron distribution, following the approach by \citet{rucinski_fahr:91}. Since the analytical formula for the electron ionization rate is still quite complex and considerably slows down the simulation, we used an approximate phenomenological formula, where the heliocentric distance $r$ is expressed in AU and the rate $\beta_{\mathrm{ el}}\left(r\right)$ at $r$ is expressed in s$^{-1}$ by:
\begin{equation}
\label{ee}
r^2 \beta _{\mathrm{ el}}(r)=\exp \left(-\frac{10.95 (\ln  r - 124.1)(\ln r +6.108)}{(\ln  r -7.491)(65.25 + \ln r (\ln r + 15.63))}\right)
\end{equation}
The formula, shown in Fig.\ref{xj}, is valid to the distance of $\sim 10$~AU, beyond which the electrons are cooled so much that their ionization capability is negligible. 

Observations done with Ulysses \citep{phillips_etal:95b, issautier_etal:98} suggest that the electron ionization rate is a 3D, time dependent function of the solar cycle phase. \citet{mcmullin_etal:04a} came to a similar conclusion for the electron impact rate of helium. However, the PUI measurements with Ulysses were collected outside $\sim 2$~AU, where the relative contribution of electron ionization is already very small. Therefore, the simplified model presented above is well justified. A model of the electron-impact ionization rate relevant for lower heliocentric distances, including its latitudinal evolution during the solar cycle, has recently been presented by \citet{bzowski:08a}.

\subsubsection{Net ionization rate}

\begin{figure}
\centering
\includegraphics[width=8cm]{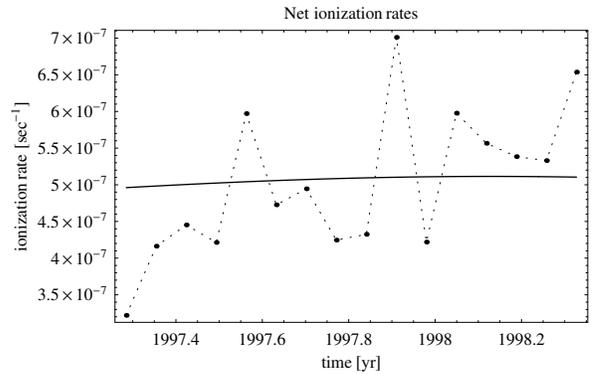}
\caption{Net ionization rates scaled to 1~AU as a function of time. Dots represent the ``instantaneous'' model, the solid line is the ``smooth'' model evaluated for 0\degr heliolatitude. }
\label{xn}
\end{figure}

\begin{figure}
\centering
\includegraphics[width=8cm]{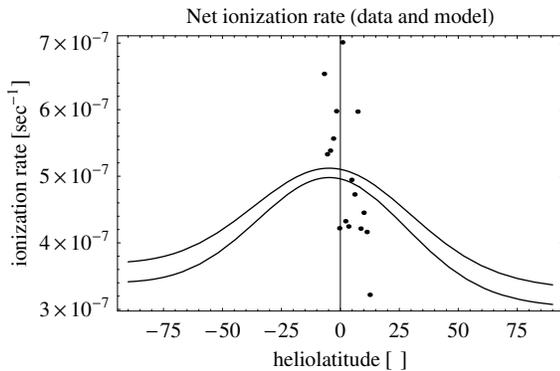}

\caption{Net ionization rates scaled to 1~AU as a function of heliolatitude. Dots represent ``instantaneous'' values and solid lines the 3D ``smooth'' evaluated for the beginning and end of the observation interval.}
\label{xl}
\end{figure}

The net ionization rates were calculated as a sum of the charge exchange, photoionization, and electron impact rates. They are presented in Fig \ref{xn} as a function of time, normalized to the latitude of ecliptic plane at 1~AU, and in Fig. \ref{xl} as a function of heliolatitude, both for the ``instantaneous'' and ``smooth'' models. In the simulations we used (i) a spherically symmetric, instantaneous model, (ii) a 3D (anisotropic) instantaneous model, (3) a spherically symmetric smooth model, and (4) a 3D smooth model. 

\end{appendix}

\begin{acknowledgements}
This research was performed within the framework of an International Space Science Institute (Bern, Switzerland) Working Group {\em Neutral Interstellar Hydrogen}. The hospitality of the ISSI Institute and the friendliness of its staff is greatly appreciated. The authors are grateful for helpful discussions with Martin Lee during the manuscript preparation. The SOLAR2000 Research Grade historical irradiances are provided courtesy of W.~Kent Tobiska and SpaceWx.com. These historical irradiances have been developed with funding from the NASA UARS, TIMED, and SOHO missions. The OMNI data were obtained from the GSFC/SPDF OMNIWeb interface at http://omniweb.gsfc.nasa.gov. The SRC PAS portion of this research was supported by the Polish MSRiT grants 1P03D00927 and N 522~022~31/0902. V.I. was supported in part by RFBR grants 07-02-01101-a, 07-01-00291-a and Dynastia Foundation. Support for this study from NASA Grant NNG06GD55G and Grant NAG 5-12929 through a subcontract from the California Institute of Technology is gratefully acknowledged.
\end{acknowledgements}

\bibliographystyle{aa}
\bibliography{iplbib}
\end{document}